\documentclass[twocolumn,prx,showkeys,superscriptaddress,longbibliography,nobibnotes]{revtex4-1}

\usepackage{graphicx,float,chemformula}

\usepackage{siunitx}

\usepackage{amsmath,amssymb,amsfonts,dsfont,bbm,mathrsfs,bm,
braket,color,graphicx,comment,extarrows}

\usepackage{lipsum}
\usepackage[colorlinks,citecolor=blue,urlcolor=blue]{hyperref}

\definecolor{purple}{rgb}{0.7, 0., 0.8}
\definecolor{grun}{rgb}{0.0, 0.7, 0.0}
\definecolor{hunoz}{rgb}{1, 0.5, 0.0}
\begin{document}

\title{Observation of an unexpected negative magnetoresistance in magnetic Weyl semimetal \ch{Co3Sn2S2}}

\author{Ali G. Moghaddam}\email{agorbanz@iasbs.ac.ir}
\affiliation{Leibniz Institute of Solid State and Materials Science (IFW Dresden), Helmholtzstr. 20, 01069 Dresden, Germany}
\affiliation{Department of Physics, Institute for Advanced Studies in Basic Sciences (IASBS), Zanjan 45137-66731, Iran}
\affiliation{Computational Physics Laboratory, Physics Unit, Faculty of Engineering and
Natural Sciences, Tampere University, FI-33014 Tampere, Finland}

\author{Kevin Geishendorf}
\affiliation{Leibniz Institute of Solid State and Materials Science (IFW Dresden), Helmholtzstr. 20, 01069 Dresden, Germany}

\author{Richard Schlitz}
\affiliation{Institut fuer Festkoerper- und Materialphysik, Technische Universitaet Dresden}
\affiliation{Department of Materials, ETH Zürich, 8093 Zürich, Switzerland}

\author{Jorge I. Facio}
\affiliation{Leibniz Institute of Solid State and Materials Science (IFW Dresden), Helmholtzstr. 20, 01069 Dresden, Germany}
\affiliation{Centro At\'omico Bariloche, Instituto de Nanociencia y Nanotecnolog\'ia (CNEA-CONICET) and Instituto Balseiro. Av. Bustillo 9500, Bariloche (8400), Argentina}

\author{Praveen Vir}
\affiliation{Max Planck Institute for Chemical Physics of Solids, 01187 Dresden, Germany}

\author{Chandra Shekhar}
\affiliation{Max Planck Institute for Chemical Physics of Solids, 01187 Dresden, Germany}

\author{Claudia Felser}
\affiliation{Max Planck Institute for Chemical Physics of Solids, 01187 Dresden, Germany}

\author{Kornelius Nielsch}
\affiliation{Leibniz Institute of Solid State and Materials Science (IFW Dresden), Helmholtzstr. 20, 01069 Dresden, Germany}
\affiliation{Technische Universität Dresden, Institute of Applied Physics, D-01062 Dresden, Germany}
\affiliation{Technische Universität Dresden, Institute of Materials Science, D-01062 Dresden, Germany}

\author{Sebastian T. B. Goennenwein}
\affiliation{Fachbereich Physik, Universität Konstanz, D-78457 Konstanz, Germany}

\author{Jeroen van den Brink}
\affiliation{Leibniz Institute of Solid State and Materials Science (IFW Dresden), Helmholtzstr. 20, 01069 Dresden, Germany}
\affiliation{Institute for Theoretical Physics and Würzburg-Dresden Cluster of Excellence ct.qmat, Technische Universität Dresden, 01069 Dresden, Germany}

\author{Andy Thomas}\email{andy.thomas@tu-dresden.de}
\affiliation{Leibniz Institute of Solid State and Materials Science (IFW Dresden), Helmholtzstr. 20, 01069 Dresden, Germany}
\affiliation{Technische Universität Dresden, Institut für Festkörper- und Materialphysik, 01062 Dresden, Germany}

\begin{abstract}
Time-reversal symmetry breaking allows for a rich set of magneto-transport properties related to electronic topology.
Focusing on the magnetic Weyl semimetal \ch{Co3Sn2S2}, we prepared micro-ribbons and investigated their transverse and longitudinal transport properties from \SIrange{100}{180}{K} in magnetic fields $\mu_0 H$ up to \SI{2}{T}. We establish the presence of a magnetoresistance (MR) up to \SI{1}{\percent} with a strong anisotropy depending the projection of $H$ on the easy-axis magnetization, which exceeds all other magnetoresistive effects.
Based on detailed phenomenological modeling, we attribute the observed results with unexpected form of anisotropy 
to magnon MR resulting from magnon-electron coupling.
Moreover, a similar angular dependence is also found in the transverse resistivity which we show to originate from the combination of ordinary Hall and anomalous Hall effects. 
Thus the interplay of magnetic and topological properties governs the magnetotransport features of this magnetic Weyl system.
\end{abstract}

\keywords{Magnetic Weyl Semimetals, Transport properties, Anisotropic magnetoresistance, Anomalous Hall effect, Electron-magnon coupling}

\maketitle

\section{Introduction}
Over the past decade, prediction and experimental realization of topological semimetals including Dirac and Weyl semimetals (WSMs) have opened a new era in condensed matter physics from both fundamental and practical points of views \cite{Burkov2011,Wan2011,Yan2017,Armitage2018,Xu2015,Xu2015b,Lv2015,Lv2015b,Yang2015,Huang2015,Xue2015TaP,xu2016observation,Souma2016NbP,Gooth2017}. Most of the intriguing properties of WSMs stem from topologically-protected band crossings known as Weyl nodes which act as Berry curvature monopoles \cite{fang2003anomalous,Armitage2018}. The nontrivial band structure manifests in a variety of physical phenomena such as Fermi-arc surface states, 
large magnetoresistance and intrinsic anomalous Hall effect (AHE) \cite{spivak2013,burkov2015,xiong2015evidence,zhang2016signatures,hirschberger2016chiral,huang2015chiral,Reis2016,ong2018,burkov2017,tewari2017,wu2018planar,kumar2018}.
The existence of isolated non-degenerate Weyl nodes requires breaking of either time-reversal or inversion symmetry \cite{Armitage2018}. While the first experimentally realized topological semimetals belong to the first category \cite{Xu2015,Xu2015b,Lv2015,Lv2015b,Yang2015,Huang2015,Xue2015TaP,xu2016observation,Souma2016NbP},
 recently, a few magnetic WSMs with broken time-reversal symmetry (TRS) have become available, opening the possibility to study the interplay between magnetism and nontrivial electronic topology. 

\begin{figure}
\includegraphics[width=\linewidth]{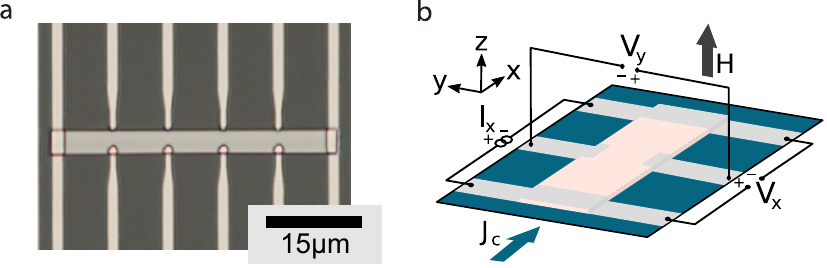}
\caption{(a) Optical micrograph of the device consisting of a \SI{60}{\micro m} long \ch{Co3Sn2S2} nanoribbon in the center and the Ti/Pt contacts around it. (b) Contact geometry for the magnetoresistance measurements, i.e.\ longitudinal ($V_x$) and transverse ($V_y$) voltage in dependence of the current ($I_x$) and the external magnetic field ($H$).}
\label{overview}
\end{figure}

\begin{figure*}
\includegraphics[width=0.8\linewidth]{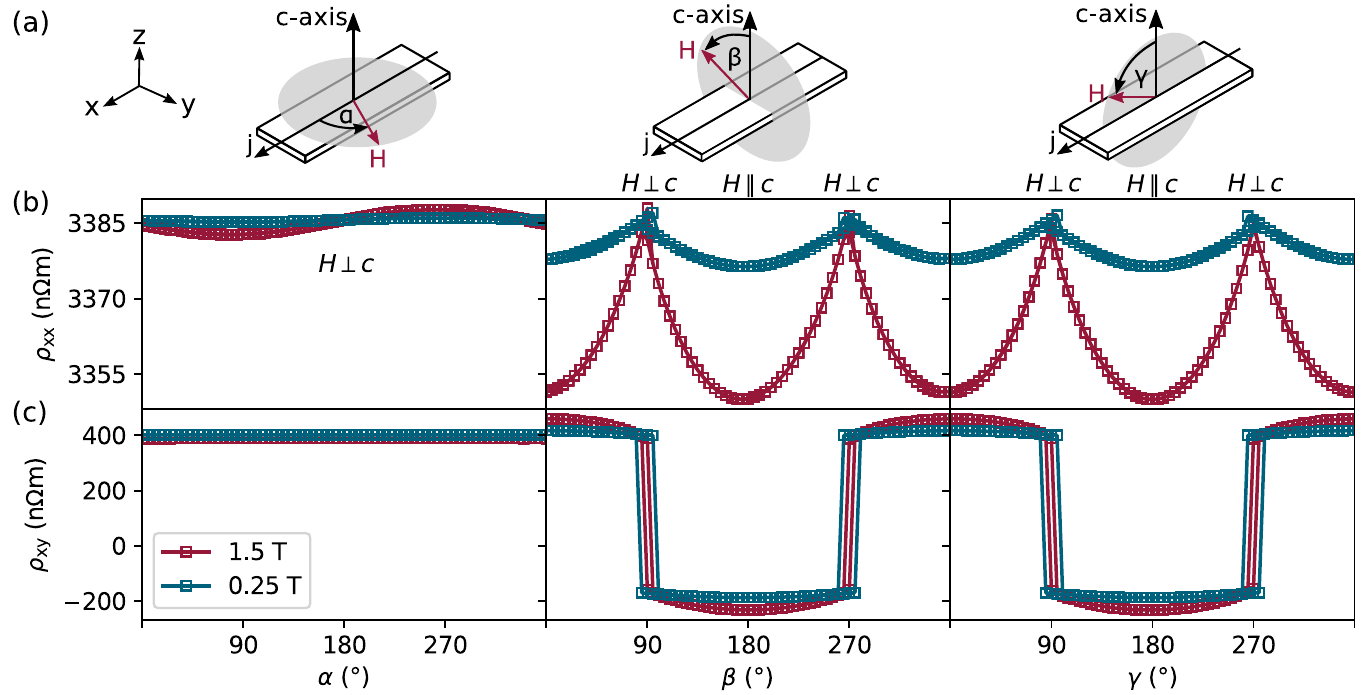}
\caption{
Longitudinal (b) and transverse (c) magnetoresistance of the \SI{40}{\micro m} long \ch{Co3Sn2S2} nanoribbon in dependence of the magnetic field orientation for rotations of $H$ around the  $z$, $x$, and $y$ axis, respectively, as indicated in (a). The data were taken at \SI{170}{K}.}
\label{fig:DMR}
\end{figure*}

One of the most appealing examples of WSM with broken
TRS is the shandite-type Kagome-lattice \ch{Co3Sn2S2}, which exhibits a strong uniaxial magnetocrystalline anisotropy \cite{Schnelle2013,Kassem2016JPSJ,Kassem:2017cy} with easy-axis ferromagnetic ordering \cite{liu2018giant,Wang2018Co3Sn2S2}. 
The topological nature of this material as a time-reversal-breaking WSM has been recently justified by visualizing its characteristic surface Fermi-arcs in different measurement setups \cite{liu2019magnetic,morali2019fermi}.
It also exhibits a large AHE \cite{Wang2018Co3Sn2S2,Geishendorf:2019iv} and an anomalous Nernst effect with promising applications for spintronic and thermoelectric devices \cite{guin2019zero,PhysRevX.9.041061,geishendorf2019signatures,Fu2020topothermo}. Here, using a micro-ribbon specimen of single-crystalline \ch{Co3Sn2S2} as depicted in Fig. \ref{overview}, we measure the transverse and longitudinal transport properties for a range of temperatures and in (rotating) magnetic fields $\mu_0 H$ of up to \SI{2}{T}. We find a significant MR on the order of 1\% with a particular anisotropy: The MR only depends on the relative orientation between the applied field and the easy-magnetization $c$-axis. 
We relate this anisotropic MR to magnon-electron coupling using a phenomenological model which effectively describes the measurements for temperatures above \SI{100}{K}, in both ferromagnetic and paramagnetic phases of the material.

\section{Experimental setup and methods}
\ch{Co3Sn2S2} exhibits a rhombohedral crystal structure (hexagonal setting, space group (R\={3}m) \cite{Schnelle2013} and is a member of the shandite family, where the cobalt atoms form a kagome lattice. It has a Curie temperature of $T_c=$\SI{175}{K} \cite{Kassem:2017cy} and, of particular interest for our study, a large magnetic uniaxial anisotropy with the easy axis along the $c$-axis \cite{Schnelle2013,Kassem2016JPSJ,Kassem:2017cy}.
Rotating the magnetization away from the easy axis can require up to \SI{26}{T} at very low temperatures \cite{PhysRevResearch.1.032044}. 
Close to $T_c$, a temperature difference of \SI{10}{K} induces a change from a hard-magnetic ferromagnet at \SI{170}{K} to a paramagnet at \SI{180}{K}. 

We used focused ion beam (FIB) cutting to prepare micro-ribbons out of bulk single crystals. This allows to prepare samples such that the transport response along particular crystal axes can be straightforwardly probed. In previous work, we have shown that the FIB cut single crystals exhibit the same properties as their bulk counterparts \cite{Geishendorf:2019iv}. After the ribbon was cut, the sample was transferred on a glass substrate using a micromanipulator. Details of this technique are reported by Overwijk et al.\ \cite{Overwijk:1998co}. Then, optical lithography and a lift-off process were utilized to define contact pads. \SI{10}{nm} Ti and \SI{150}{nm} Pt were deposited after a \SI{2}{min} in-situ cleaning of the contact areas. This leads to a sample structure as depicted in Fig. \ref{overview}(a). The device layout is designed in classic Hall-bar geometry, i.e.\ longitudinal and transverse sample contacts. A cartoon of the measurement setup is shown in Fig \ref{overview}(b). The device is finally mounted onto a chip carrier and wire bonded for the electrical contacts to the magnet cryostat probe. The transport measurements are carried out utilizing a variable-temperature inset and a superconducting solenoid magnet. We here focus on a device made from an approximately \SI{40}{\micro m} long,  \SI{4}{\micro m} wide and  \SI{0.35}{\micro m} thick micro-ribbon, cut such that the c-axis (the magnetic easy axis) is oriented 'out-of-plane' along $z$.

\begin{figure*}
\includegraphics[width=0.8\linewidth]{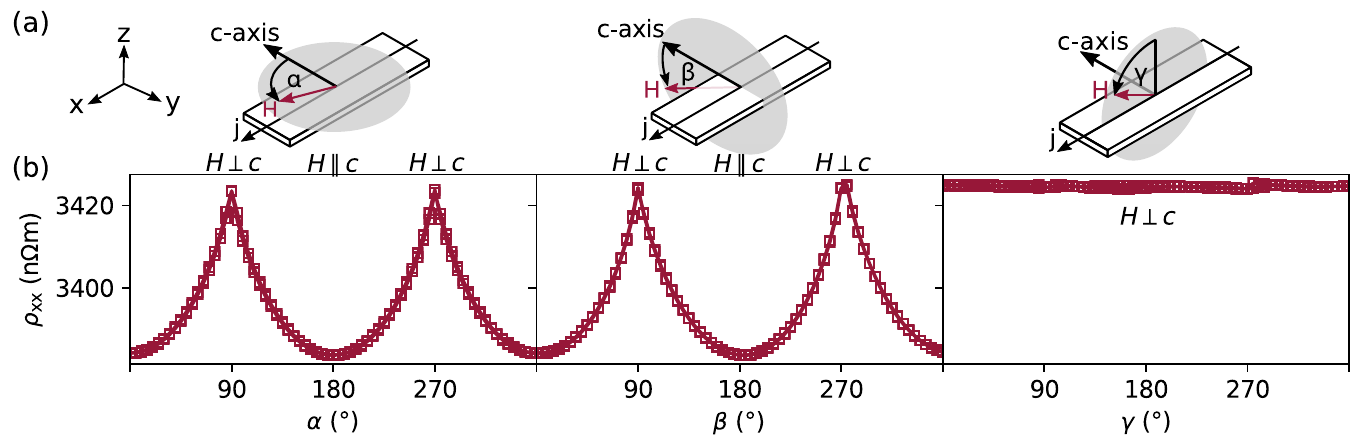}
\caption{Longitudinal magnetoresistance (b) in dependence of the angle in the three orthogonal rotation planes as indicated in (a). Please note that the $c$-axis was rotated by 90 degrees in the FIB-cut of this micro-ribbon when compared to Fig. \ref{fig:DMR}.}
\label{secondCut}
\end{figure*}

We first present magnetotransport results obtained by rotating an external magnetic field of constant magnitude in three different planes, as displayed schematically in Fig.\ \ref{fig:DMR}(a).
More specifically, we applied a constant current to the sample and recorded both the ensuing longitudinal and transverse voltages. The corresponding longitudinal ($\rho_{xx}$) and transverse ($\rho_{xy}$) resistivities for $T=$\SI{170}{K} are shown  in panels b and c, respectively, for two different magnetic field magnitudes.

The longitudinal signal exhibits a very particular form of negative MR with angular dependence as 
$\rho_{xx} = \rho_{xx}^0 - \Delta \rho_{xx} \: |\cos\varphi_{c,H}|$ 
where $\varphi_{c,H}$ denotes the angle between the $c$-axis and the external magnetic field.
Overall, an anisotropic MR of $ \Delta \rho_{xx}/\rho_{xx} =1.5$\% is observed in the presence of an external field $\mu_0 H=\SI{1.5}{T}$. 
The transverse signal, which is due to the anomalous Hall effect, shows a switching between two levels, justified by the strong uniaxial anisotropy.
The magnetization
is always parallel to $c$-axis (${\bf M}\parallel c$), but switches its sign when $\varphi_{c,H}>\pi/2$, which leads to the same behavior for 
the transverse resistivity. 
In fact, $\varphi_{M,H}$, the angle between the magnetization and magnetic field 
always remains smaller than $\pi/2$ and therefore, we have $\cos\varphi_{M,H}=|\cos\varphi_{c,H}|$.

As it can be seen in Fig.\ \ref{fig:DMR}(c), the transverse signal is not fully antisymmetric under $\beta\to180^{\circ}+\beta$. There are various possible mechanisms for such an offset in the transverse resistivity. A common source is due to the non-ideal placement of contacts, which can result in a finite percentage of longitudinal resistance in the transverse signal. In addition, it has been recently shown that the band structure of this compound is not symmetric between parallel and antiparallel configurations of M with respect to the c-axis, a property which has been reflected in quantum Shubnikov-de Haas oscillation \cite{YeFacio2022}. Consequently, we may expect an additional asymmetry of the transverse resistivity between the parallel and antiparallel orientation
of the magnetization with the c-axis ($\beta=0^{\circ}$ versus $\beta=180^{\circ}$).

To further understand the angular dependence of the MR in our setup, we took advantage of the flexibility provided by the FIB cutting and prepared another micro-ribbon out of the same \ch{Co3Sn2S2} single crystal with a different orientation of the $c$-axis, now 
`in plane' along $y$.
The MR angular variations of this micro-ribbon are shown in Fig. \ref{secondCut}.
The results are again consistent with $(\rho_{xx}-\rho_{xx}^0) \propto -|\cos\varphi_{c,H}|$. As already discussed, the magnetization $M$ is always oriented parallel to the $c$-axis, because of the large crystalline anisotropy of \ch{Co3Sn2S2}. Consequently, the assumption of $(\rho_{xx}-\rho_{xx}^0) \propto -|\cos\varphi_{c,H}|$ is equivalent to $(\rho_{xx}-\rho_{xx}^0) \propto -\cos\varphi_{M,H}$.

\begin{figure}
\includegraphics[width=0.99\linewidth]{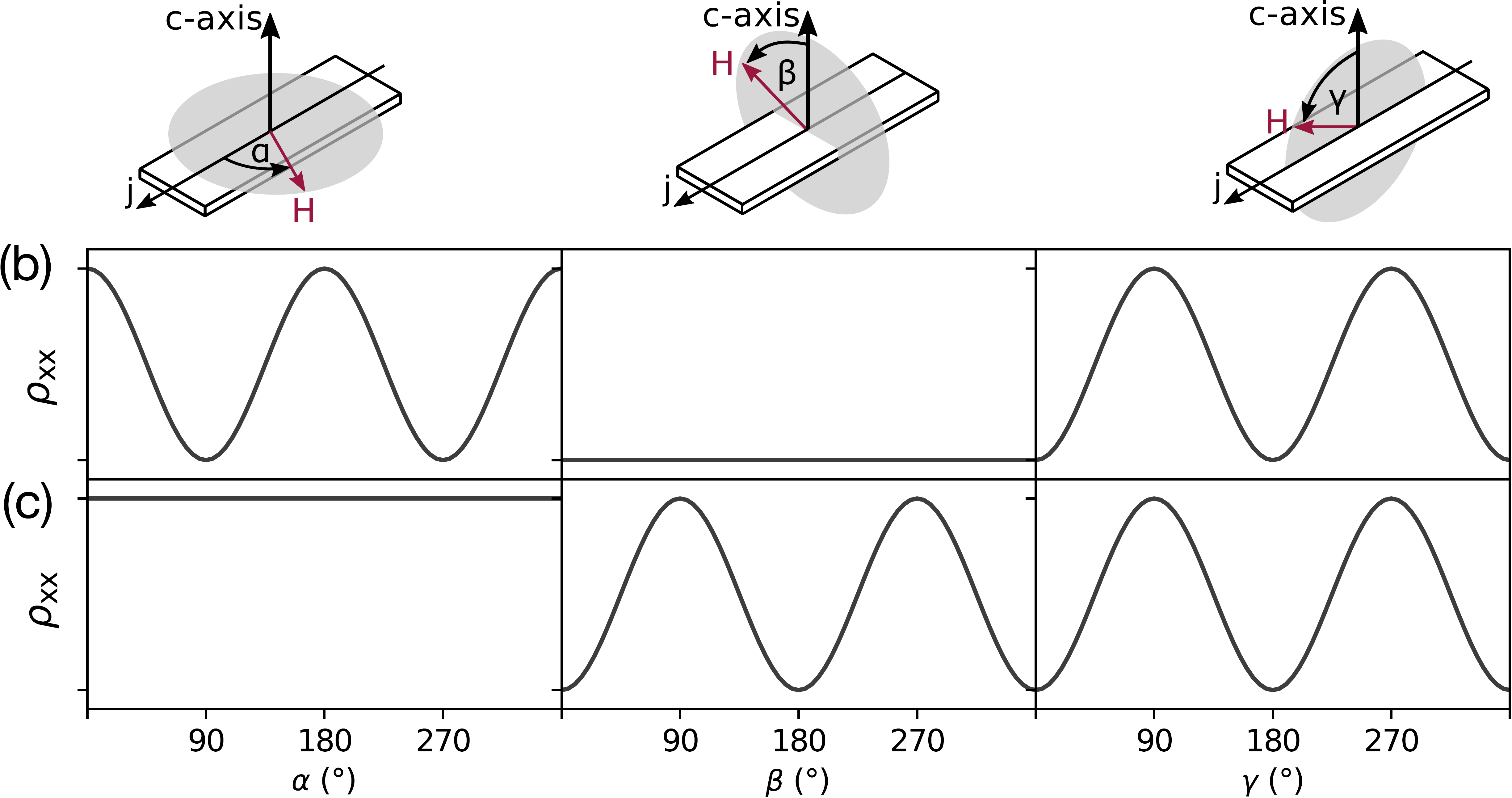}
\caption{
The three orthogonal rotation planes of the external magnetic field utilized in our experiments. (b) Conventional Lorentz deflection induced AMR with $\varphi_{j,H}$-dependence (c) Expected behavior of a magnetoresistance depending on the relative orientation of the magnetic easy axis ($c$-axis) and the external field (similar to those in Fig. \ref{fig:DMR}).
Please note that in the current schematic figure, we have simply considered $\cos\varphi$ angular dependences, for the sake of argument. 
}
\label{fig:compare}
\end{figure}

Typically, the anisotropic magnetoresistance (AMR) is governed by either the relative angle between the current and  the magnetization or the relative angle between the current and the applied magnetic field \cite{mcguire1975anisotropic}. In the first case, it can be expressed as $\rho_{xx}=\rho_{xx}^0  + \Delta \rho_{xx} \: \cos^2\varphi_{j,M}$ and one mechanism is the anisotropy of the scattering caused by the spin-orbit interactions, as first suggested by Smit \cite{smit1951magnetoresistance}. In the second case, the MR obeys $\rho_{xx}=\rho_{xx}^0  + \Delta \rho_{xx} \: \cos^2\varphi_{j,H}$ and originates in the Lorentz deflection of charge carriers due to the perpendicular component of the external field.
In our device at a relatively high temperature ($T=$\SI{170}{K}), the observed behavior is quite different from both these two angular dependencies.

For the sake of clarity, we will distinguish between conventional AMR as described in the previous paragraphs and the more general term \emph{anisotropic} MR, i.e.\ any MR exhibiting anisotropy. Intriguingly, the conventional AMR is almost absent in our measurements performed close to the magnetic transition temperature.
This can be quite immediately understood by comparing the angular variations for a conventional AMR
which only depends on $\varphi_{j,H}$ with those for an AMR 
which only depends on $\varphi_{c,H}$, as depicted in Fig. \ref{fig:compare}. Provided that our experimental results show profound dependence only on the relative angle 
$\varphi_{c,H}$ without current flow direction involved, the conventional AMR scenarios with either
$\varphi_{j,H}$ or $\varphi_{j,M}$ dependences
are unambiguously ruled out here. 
Furthermore, our observation are far from 
the chiral transport feature of WSMs
where a negative MR depending on the relative orientation of electric and magnetic fields (${\bf E}$ and ${\bf H}$) has been observed \cite{zhang2016signatures}.
Since negative magnetoresistance has repeatedly been invoked as a
signature of the chiral anomaly, our findings represent an important 
alternative approach for the interpretation of magnetotransport effects 
in Weyl semimetals.
In the next part, we will provide a theoretical analysis
suggesting magnon-electron scattering as the likely main source of observed anisotropic MR in our experiment.
Before that, we should point out that as we see a very weak modulation of $\rho_{xx}$ during the rotation of the external field around the $c$-axis 
as shown in the rightmost panel of Fig. \ref{secondCut}(b).
These variations as a function of angle $\gamma$ in Fig. \ref{secondCut}(b) are much weaker compared to the cases where the field is swept in the two other planes corresponding to varying angles $\alpha$ and $\beta$. There is no interesting physics behind the very slight variations with the angle $\gamma$: They are mainly attributed to the a small misalignment of the rotation plane with respect to the sample.

\section{Theory of the observed anisotropic MR}
To understand the
special angular variation of $\rho_{xx}$ and $\rho_{xy}$ observed here, we present a phenomenological theory based on a two-current model for transport in ferromagnetic systems. In this framework,
the longitudinal conductivity reads \cite{Fert1969,Fert1976}
\begin{equation}\label{2-channel-mixing}
\sigma_{xx}=\frac{\rho_\uparrow+\rho_\downarrow+4\rho_m}{\rho_\uparrow\rho_\downarrow+\rho_m\,(\rho_\uparrow+\rho_\downarrow)},
\end{equation}
in terms of the two spin channels resistivities $\rho_\sigma$
and the mixing resistivity $\rho_m$.
In a simple version of this model, $\rho_\sigma\sim \frac{1}{3} e^2 \,v_F^2 \, \tau_\sigma \, D_\sigma(\varepsilon_F)$ 
is given by the spin-dependent density of states (DOS) $D_\sigma$ and relaxation times $\tau_\sigma$. 
In fact, the spin-dependent resistivities $\rho_\uparrow$ and $\rho_\downarrow$ can be quite different particularly in the presence of the polarization of conduction electrons defined as $P = (D_{\uparrow}-D_{\downarrow})/(D_{\uparrow}-D_{\downarrow})$.
Previous ab initio calculations have suggested that the band structure of \ch{Co3Sn2S2} is highly polarized \cite{liu2018giant}. Therefore, one may expect 
$\rho_\uparrow$ and $\rho_\downarrow$ to be significantly different in this material. Nevertheless, since the polarization of electrons is not influenced by the applied field (in the low field regime), it does not take any direct important role in the field-dependent transport properties. The term $\rho_m$ mixes the contributions of up and down spin currents and, therefore, depends on the spin-flip relaxation time governed by the not spin-conserving scatterings. At relatively high temperature, a significant contribution to $\rho_m$ comes from the magnon-electron scattering which results in a linear dependence of the resistivity on the applied magnetic fields \cite{Raquet2002}, known as \emph{magnon magnetoresistance} (MMR). As it will be become clear in the following, it governs the 
field-dependent transport properties and especially the observed anisotropic MR which can be put in the following phenomenological form
\begin{equation}
	\rho_{xx}=\rho_{xx}^0   - \eta\: {\bf M} \cdot {\bf H}
	\label{eq:mmr}
\end{equation}
where $\rho_{xx}^0 $ and $\eta$ are two constants denoting the base resistance of the micro ribbon and the strength of the MR, respectively. 

To understand the field dependence of the anisotropic MR (which is caused by magnons and for that reason can be equally called as MMR), we follow the model of Ref. \cite{Raquet2002}, where
the effect is related to the change of magnon-electron scattering through changes in the magnon population.
Based on this model, it has been already suggested that in materials with anisotropy field much larger than the applied field ($H_A\gg H$), and for collinear alignments of the field and the magnetization, 
the MMR linearly varies as $\rho_m-\rho_m^0\propto -H\:M$ 
with $\rho_m^0$ denoting the zero field value
\cite{Attane2008,Attane2011,Nguyen2011}. The underlying reason for this dependence is that the gap in the magnon dispersion 
is proportional to the total magnetic induction \cite{Taylor1968}. Therefore,
for parallel applied field and magnetization, the magnon gap  
slightly increases which in turn leads to the attenuation of the magnon population, and in turn of the electron-magnon scattering, resulting in a resistivity decrease.  
We generalize this picture to noncollinear alignments as 
$\rho_m =\rho_m^0 - \delta\rho_m |\cos\varphi_{c,H}|$ with $\delta\rho_m$ linearly dependent on $H$.
Plugging this ansatz for $\rho_m$ in Eq.
\eqref{2-channel-mixing} and assuming $\delta\rho_m\ll \rho_m^0$, we find 
\begin{align}
\label{eq:sig_xx}
\sigma_{xx}&=\sigma_{xx}^0  + \delta\sigma_{xx} |\cos\varphi_{c,H}|,
\end{align}
where $\delta\sigma_{xx}$ again is proportional to the field strength $H$.

Besides the longitudinal conductivity, we have a transverse conductivity $\sigma_{xy}$ attributed to the AHE consisting of intrinsic (Berry curvature) and extrinsic 
(skew scatterings and side jump)
contributions \cite{nagaosa2006,Nagaosa2010}. 
The skew scattering contribution which is linear in the magnetization $M_z$ \cite{zeng2006linear}, becomes dominant in \ch{Co3Sn2S2} at higher temperatures \cite{parkin2020nanolett}.
Now, in the presence of an applied magnetic field $\sigma_{xy}$ has an additional contribution from the ordinary Hall effect (OHE) as $\sigma_{xy}=\sigma_{xy}^{\rm AHE}(M_z)  + \sigma_{xy}^{\rm OHE}(H_z)$ which can be recast into
\begin{align}
\label{eq:sig_xy}
\sigma_{xy}&=\sigma_{xy}^0\:{\rm sgn}(M_z) + \delta\sigma_{xy} \cos\varphi_{c,H},
\end{align}
assuming only the sign of magnetization can change. In the above expression, $\sigma_{xy}^0$ accounts for the pure AHE in the absence of an external field, 
whereas the second term which comes from the linear dependence on the perpendicular field $H_z = H \cos\varphi_{c,H}$, can be attributed to the OHE and the slight change of the magnetization due to the external field. Note that disentangling the contribution of ordinary and anomalous Hall effects is only possible due to the strong magnetic anisotropy of this material.

Assuming weak angular dependence in Eqs. \eqref{eq:sig_xx} and \eqref{eq:sig_xy} meaning that $\delta\sigma_{ij}/\sigma^0_{ij} \ll 1$ as suggested by the experimental results, the corresponding first order corrections to the resistivities are found to be
\begin{align}
\label{eq:rho_xx}
\rho_{xx}&\approx\rho_{xx}^0\: \big(1 - \epsilon_{xx}  |\cos\varphi_{c,H}|  \big), \\
\label{eq:rho_xy}
\rho_{xy}&\approx\rho_{xy}^0 \: \big[{\rm sgn}(M_z) + \epsilon_{xy}  \cos\varphi_{c,H}
\big].
\end{align}
Here, $\rho_{ij}^0$ denotes the zero-field resistivities, and dimensionless prefactors  
$\epsilon_{ij}=\delta\sigma_{ij} /\sigma_{ij}^0 $ ($i,j=x,y$)
represent the field-dependent coefficients. From the measurement results at the field 
\SI{1.5}{T} shown in Fig. \ref{fig:DMR}, we
estimate the coefficients to be $\epsilon_{xx}\sim 10^{-2}$ and $\epsilon_{xy}\sim 10^{-1}$ giving rise to a very good quantitative agreement between the measurements and the theoretical model. This justifies the assumption leading to above equations, whereby higher-order corrections showing up by inverting the conductivity tensor can be safely dropped. We also remind that Eq. \eqref{eq:rho_xx} is fully consistent with \eqref{eq:mmr}, considering $\eta M H \equiv \rho^0_{xx}\epsilon_{xx}$.

\begin{figure}
\includegraphics[width=\linewidth]{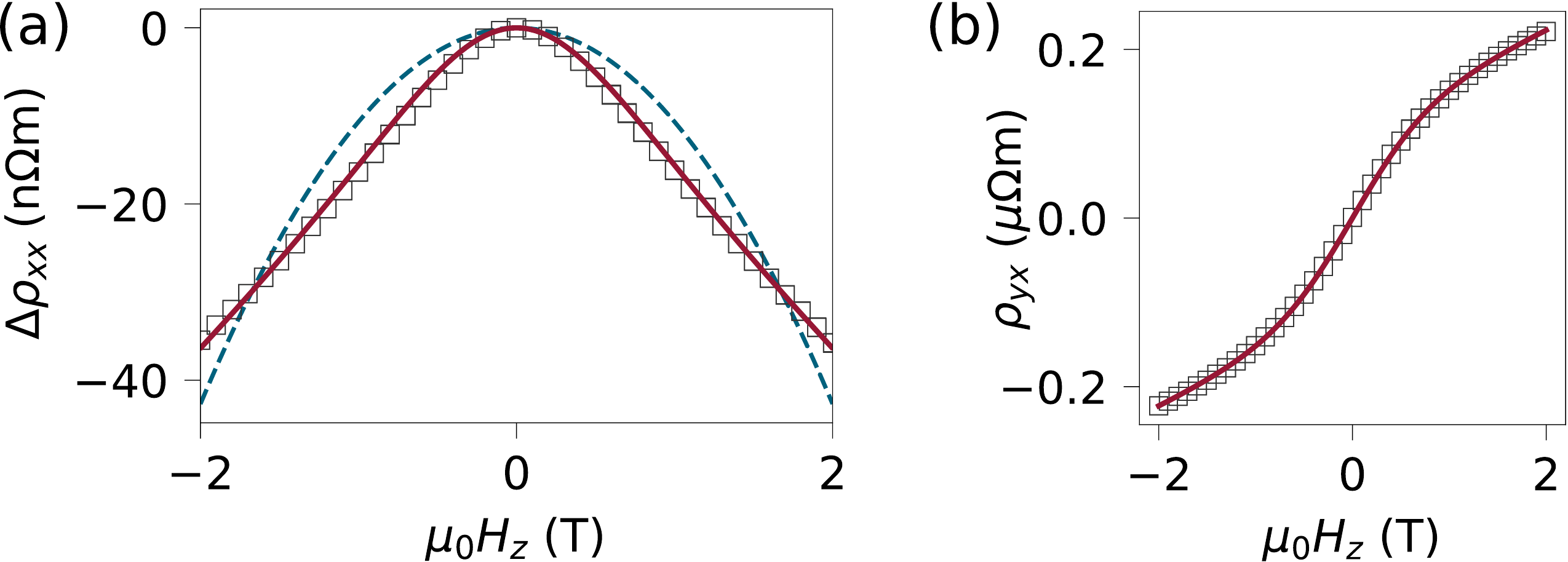}
\caption{ (a) Field-dependent longitudinal resistivity $\Delta\rho_{xx}$ at \SI{180}{K} for external magnetic fields applied along the $z$-direction (open symbols). The dashed blue line represents a fit to $\Delta\rho_{xx} \propto H^2$. The solid red line represents a calculation using $\Delta\rho_{xx} \propto H \cdot M$.
(b) Transverse resistance (open symbols) corresponding to the AHE and Langevin fit of these values (red line).
}
\label{closerLook1}
\end{figure}

\section{MR in the paramagnetic phase}
In the paramagnetic phase, the spontaneous magnetization vanishes, yet in the presence of an external magnetic field, a finite magnetization is induced as $M(H) = \chi(H) H$ \cite{ohandley2000}, where $\chi(H)$ denotes the angle dependent magnetic susceptibility. Using the AMR relation (\ref{eq:mmr}) caused by magnons, we expect the field-dependent contribution of the longitudinal resistivity denoted by $\Delta\rho_{xx}=\rho_{xx}-\rho^0_{xx}$ to be proportional to $MH$, and subsequently we have $\Delta\rho_{xx} \propto \chi(H) H^2$ as $M$ and $H$ are parallel to each other. To check the validity of this picture, we performed an additional set of measurements of longitudinal resistivity at $T=$\SI{180}{K}.
In this case, the field is not rotated, but swept along the the direction of magnetic easy axis. This is depicted in Fig. \ref{closerLook1}(a) (open symbols).

We extract $\chi(H)$ from the AHE and the corresponding transverse resistivity $\rho_{xy}$ shown in Fig. \ref{closerLook1}(b) via
\begin{equation}
    \rho_{xy}=R_H H + R_{AH}M(H)=\big[R_H + R_{AH}\chi(H)\big]H, \label{eq:rho_xy-para}
\end{equation}
with ordinary Hall and AHE coefficients $R_H$ and $R_{AH}$, respectively. The magnetization is expected to match the Langevin function $L(\xi)=\coth(\xi)-1/\xi$ as $M(H)/M_0=L(\mu H/k_BT)$, since the temperature of \SI{180}{K} is above the ferromagnetic ordering temperature of \ch{Co3Sn2S2} (\SI{175}{K}). So by fitting $\rho_{xy}$ with the expression \eqref{eq:rho_xy-para} shown by a red line in Fig. \ref{closerLook1}(b), we acquire the unknown parameters of the Langevin function i.e.\ $M_0$ and $\mu$. Using the magnetization calculated with the Langevin function allows us to  quantitatively evaluate the MMR expressed by Eq. \eqref{eq:mmr}
and the corresponding field dependent contribution to the longitudinal resistivity $\Delta\rho_{xx}$. This is indicated by the red line in Fig. \ref{closerLook1}(a) which shows a very good agreement with the experimental data (open symbols), particularly in contrast to the the naive quadratic approximation (blue dashed line) as previously discussed \cite{doi:10.1143/JPSJ.34.51,10.1143/PTP.48.1828}.

\begin{figure}
\includegraphics[width=\linewidth]{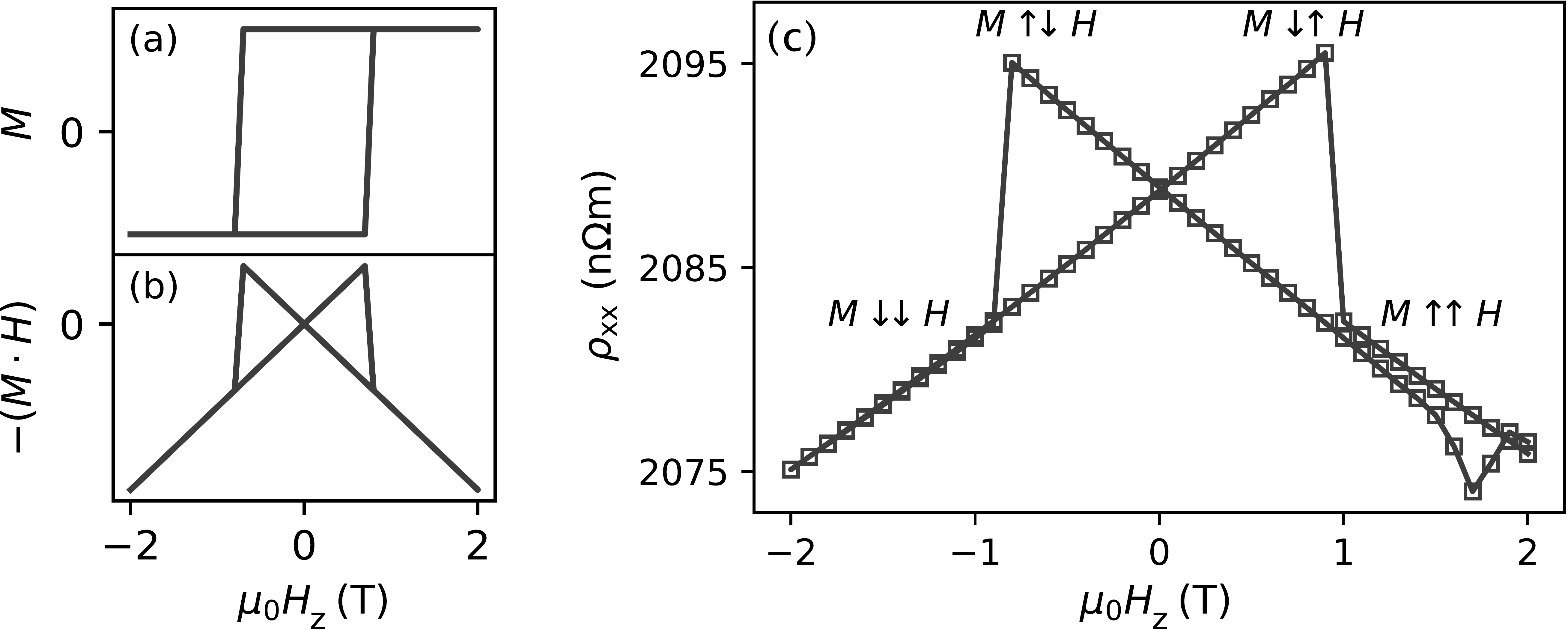}
	\caption{Schematic (calculated/ simulated) behavior of (a) $M$, (b) $M\cdot H$ for a magnetic field sweep parallel to the magnetic easy axis. (c) Experimental results for the longitudinal resistivity at \SI{100}{K}.}
\label{batman}
\end{figure}

It should be emphasized that the observed data exhibits a nonlinear field-dependence of the magnetization $M$. This behavior is expected, since we are slightly above the transition temperature. At very high temperatures and deep in the paramagnetic phase, 
the magnetic susceptibility becomes field independent, thereby, the magnetization 
approaches the linear dependence on $H$. As a result and at very high temperatures, the longitudinal and transverse resitivities should be quadratic and linear with $H$, respectively.

\section{MMR at lower temperatures}
As another check for the MMR fingerprint as given by \eqref{eq:mmr}, we also performed longitudinal resistivity $\rho_{xx}$ measurements at
an intermediate temperature \SI{100}{K} and by sweeping the magnetic field along the $c$-axis. In this case, we expect a bow-tie structure of the resulting magnetoresistance as depicted in Fig. \ref{batman}(a-b). The corresponding experimental results for $\rho_{xx}$ are shown in Fig. \ref{batman}(c). 
While the coercive field in
the microstructures is further increased by the shape anisotropy, the bow-tie shape of the measurements directly corresponds to the predicted behavior assuming MMR with linear dependence on ${\bf M}\cdot {\bf H}$.

\section{Summary}
We have explored the magnetotransport properties of 
the magnetic Weyl semimetal \ch{Co3Sn2S2} in micro-ribbon geometries and found a strong anisotropy in the longitudinal magnetoresistance as well as in the transverse resistivity. The angular dependence is governed by the angle enclosed between the external magnetic field and the magnetization. We provide a simple phenomenology for the observed anisotropic MR based 
on the contribution of electron-magnon scattering
to transport and on the interplay between ordinary and anomalous Hall effects. 
The model also successfully describes measurements both in  the paramagnetic and in the ferromagnetic phases for temperatures above \SI{100}{K}.
Our findings provide new insights into the magnetotransport features of this magnetic Weyl system mainly inherited from its strong magnetocrystalline anisotropy together with its large anomalous Hall response.

\section{Acknowledgments}
The authors would like to thank Almut P\"ohl for the preparation of the lamellae. J.I.F.\ acknowledges the support from the Alexander von Humboldt Foundation. J.v.d.B., C.F.\ and S.T.B.G.\ acknowledge financial support from the German Research Foundation (Deutsche Forschungsgemeinschaft, DFG) via SFB 1143.

\section{Author Contributions}
S.T.B.G and A.T. conceived and supervised the experiments.
K.G. and R.S. performed the sample fabrication, measurements and analysis.
P.V. and C.S. made the single crystal samples.
A.G.M., J.F., and J.v.d.B carried out the theoretical analysis and modelling.
K.N. and C.F. provided oversight and advice.
A.G.M., S.T.B.G, and A.T. wrote the manuscript and all authors contributed to its final version.

\bibliography{Co3Sn2S2rotation}

\end{document}